\documentstyle[preprint,aps,prb]{revtex}
\begin{document}
\tightenlines
\title{Ground-state energy of confined charged bosons in two 
dimensions}
\author{A. Gonzalez$^{1,2}$\cite{adres}, 
B. Partoens$^3$\cite{bart}, A. Matulis$^{3}$\cite{algis} and 
F. M. Peeters$^3$\cite{francois}}
\address{$^1$Departamento de Fisica, Universidad de Antioquia, 
AA 1226, Medellin, Colombia\\ 
$^2$Instituto de Cibernetica, Matematica y Fisica Calle E 309, 
Vedado, Habana 4, Cuba\\
$^3$Departement Natuurkunde, Universiteit Antwerpen (UIA),
Universiteitsplein 1, B - 2610 Antwerpen, Belgium}
\date{\today}

\maketitle

\begin{abstract}
The Pad\'e approximant technique and the variational Monte Carlo method
are applied to determine the ground-state
energy of a finite number of charged bosons in two dimensions confined by
a parabolic trap.
The particles interact repulsively through a Coulombic, $1/r$, potential.
Analytic expressions for the ground-state energy are obtained.
The convergence of the Pad\'e  sequence and comparison with 
the Monte Carlo results show that the error of the Pad\'e estimate 
is less than $4~\%$ at any boson density and is exact in the 
extreme situations of very dilute and high density.
\end{abstract}

\pacs{PACS numbers: 05.30.Jp, 03.75.Fi, 02.70.Lq}

The interest in finite boson systems which are confined in traps 
increased suddenly
after the experimental observation of signals of Bose-Einstein (BE)
condensation in clusters of alkali-vapour atoms.\cite{BE} These are confined
three-dimensional dilute systems interacting through short-range 
potentials, which basically can be described by a modification of 
Bogoliubov's theory.\cite{GPS97}

Lower-dimensional systems as well as bosons interacting through 
long-range potentials are also widely studied. These studies are motivated,
for example, by proposals of obtaining quasi 
two-dimensional clusters by making the traps highly
anisotropic.\cite{quasi2D} 
Theoretically some regularities of the 
spectra of energy levels have been identified to check experimentally 
the effective quasi-two-dimensionality of the system.\cite{P96} A particularly 
interesting aspect of trapped systems is related to the fact that 
quantum fluctuations do not destroy BE condensation in lower
dimensions.\cite{D<3} On the other hand, the ground-state properties of the 
homogeneous charged Bose 
gas has been studied in 3D.\cite{3D}
Vortex systems in high-$T_c$ superconductors have been shown 
to be equivalent to 2D bosons with logarithmic interactions.\cite{vortex} 

In the present paper, we study a model of two-dimensional 
charged bosons confined in a parabolic trap. Instead of an often used  
long-wavelength field theory description, we start from the 
$N$-particle hamiltonian, which in harmonic oscillator units 
(i.e. $\sqrt{\hbar/(m \omega_0)}$ for length, $\hbar \omega_0$ for
energy, etc.) reads

\begin{equation}
\frac{H}{\hbar\omega_0}=\frac{1}{2}\sum_{i=1}^N (\vec p_i^{~2}+
  \vec r_i^{~2})+\beta^3 \sum_{i<j}\frac{1}{|\vec r_i-\vec r_j|},
\label{hamiltonian}
\end{equation}

\noindent
where $\omega_0$ is the dot confinement frequency, $\beta^3=\sqrt{(\frac{m
q^4}
{\hbar^2})/(\hbar\omega_0)}$, and $m$ and $q$  are, respectively, 
the particle mass and charge. By varying $\beta$ we modify the ``density'' 
of the system.
\par
Following our approach for electrons in a parabolic quantum dot,\cite{GPP97}
we will construct two-point Pad\'e approximants for the ground-state energy
of the Hamiltonian~(\ref{hamiltonian}) from the weak and strong
interaction expansions of the energy
\begin{eqnarray}
\left. \epsilon\right|_{\beta\to 0} &=& b_0 +b_3\beta^3 +b_6\beta^6
    + \dots ,\label{hdlimit}\\
\left. \epsilon\right|_{\beta\to\infty} &=& \beta^2 \{a_0 +a_2/\beta^2
    + \dots\}\label{ldlimit}.
\end{eqnarray}

\noindent
When $\beta<<1$, we have a system of weakly interacting bosons in a 
harmonic potential. For $\beta=0$ we have $N$ non-interacting bosons with energy
$b_0=N$ which is the leading contribution to the energy~(\ref{hdlimit}).
Perturbation theory is applied to compute $b_3$ and $b_6$, 
yielding

\begin{eqnarray}
b_3&=&<0|V|0>=\frac{N(N-1)}{2} \sqrt{\frac{\pi}{2}},\\
b_6&=& \sum_{int} \frac{<0|V|int>^2}{\epsilon_0(0)-\epsilon_0(int)}
           \nonumber\\ 
     &=& c_1~N(N-1)^2 + c_2~N(N-1),\label{second}\\
c_1&=&-0.05300695,\;\;\; c_2=- 0.11982050, \nonumber
\end{eqnarray}

\noindent
where $V=\sum_{i<j} |\vec r_i-\vec r_j|^{-1}$, and $int$ is a shorthand 
notation for the harmonic oscillator intermediate states respecting
conservation
of angular momentum (equal to zero in the ground state). Due to the pair 
character of the potential, only one- and two-particle excitations are allowed. 
$\epsilon_0(int)$ is the energy of the non-interacting 
intermediate state and $\epsilon_0(0)=N$. 
We would like to notice that in order to apply the Pad\'{e} approximant technique one needs to
calculate the coefficients with rather high accuracy. This was achieved in the
following way.
The one-particle excitation part (namely the coefficient $c_1$)
was calculated from the straightforward summation of a single
sum
which converges rather fast. The convergence of the double sum in the
two-particle
excitation part, however, is slow. Therefore we used an alternative approach and
calculated the coefficient $c_2$
from the solution of the two particle 
interaction problem which is given in the Appendix.
%
\par
On the other hand, when $\beta\to\infty$, a scaling of coordinates 
$r\to\beta r$ in the Hamiltonian (1) shows \cite{GPP97,MP94,G97} that the 
potential energy (Coulomb repulsion plus parabolic confinement) behaves 
as $\beta^2$, whereas the 
kinetic energy (``fluctuations'') is of order zero in $\beta$. Consequently, in
this limit the system behaves as $N$ classical point particles and $a_0$ is 
the minimum of the classical potential energy. The next term $a_2$ takes 
account of the zero-point 
fluctuations, which is given by $a_2=\sum_{\alpha} \omega_{\alpha}/2$. 
The coefficients $a_0$ and $a_2$ for $N=2-5$ are given in Ref.~\onlinecite{G97}. They were obtained from a strong-coupling expansion 
of the exact quantum-mechanical Hamiltonian. For larger $N$, $a_0$ and
$a_2$
were 
obtained by minimising the potential energy with a combination of Monte Carlo 
and Newton methods and the normal frequencies were computed by 
solving a classical small-oscillation problem.\cite{BP94,SP95} Fitted 
expressions (as given in Ref.~\onlinecite{GPP97})
for the coefficients $a_0$ and $a_2$ for 
$6 \le N \le 210$ are
\begin{eqnarray}
a_0/N^{5/3} &=& 1.062 - 0.875/N^{1/2} - 0.185/N , \\
a_2/N^{5/4} &=& 0.573 + 0.475/N^{1/2} - 0.160/N . 
\end{eqnarray}
These expressions also have the correct $N\rightarrow\infty$ asymptotic
behavior. 
%
%
\par
Two-point Pad\'e approximants $\{P_{s,t}(\beta)\}$ are used to 
estimate the ground-state energy. They are quotients of polynomials reproducing 
the first  $s+1$ terms of the expansion~(\ref{hdlimit}) and the first $t+1$ terms of 
the $\beta\to\infty$ series~(\ref{ldlimit}). By construction, the approximants are 
asymptotically exact in both the $\beta\to 0$ and $\beta\to\infty$
limits. 
As it is common with the Pad\'{e} approximants, we shall check the convergence
of a sequence $\{P_{s,t}(\beta)\}$ running parallel to the diagonal
$\{P_{s,s}(\beta)\}$.\cite{MR86} 
In the present problem we use the off-diagonal sequence
$\{P_{K+3,K}(\beta)\}$, which is the first non-trivial sequence with no
singularities. We will show that it exhibits  
good convergence properties at 
intermediate $\beta$ values. The explicit form of the first elements of 
this sequence are the following

\begin{eqnarray}
P_{3,0}(\beta) &=& b_0 + \frac{b_3 \beta^3}{1+b_3 \beta/a_0},\\
P_{4,1}(\beta) &=& b_0 + b_3 \beta^3 \left\{1 - \frac{q_1^2 \beta^2}
                {1+q_1 \beta+q_1^2 \beta^2} \right\},\\
           q_1 &=& b_3/a_0,\nonumber\\
P_{5,2}(\beta) &=& b_0 + b_3 \beta^3 \left \{1-\frac{q_3 \beta^3}
               {1+q_1\beta+q_2\beta^2+q_3\beta^3}\right \},\\
           q_1 &=& a_0 q_2/b_3,~ q_3=b_3 q_2/a_0,\nonumber\\ 
           q_2 &=& \left(\frac{a_2-b_0}{a_0}+\frac{a_0^2}{b_3^2}
                   \right)^{-1},\nonumber\\
P_{6,3}(\beta) &=& b_0 + \frac{b_6 \beta^6}{1+q_1 \beta+\dots
+q_4\beta^4}\nonumber\\
               &&+ b_3 \beta^3 \left \{1-\frac{q_4 \beta^4}{1+q_1 \beta+\dots 
                                     +q_4\beta^4}\right \},\\
           q_2 &=& a_0 q_3/b_3,~ q_4=(b_3 q_3+b_6)/a_0, \nonumber\\
           q_1 &=& (q_3 (b_0-a_2)+b_3)/a_0, \nonumber\\
           q_3 &=& \frac{(a_2-b_0)|b_6|/a_0+b_3^2/a_0}
                        {a_0^2/b_3+2 b_3 (a_2-b_0)/a_0}.\nonumber
\end{eqnarray}

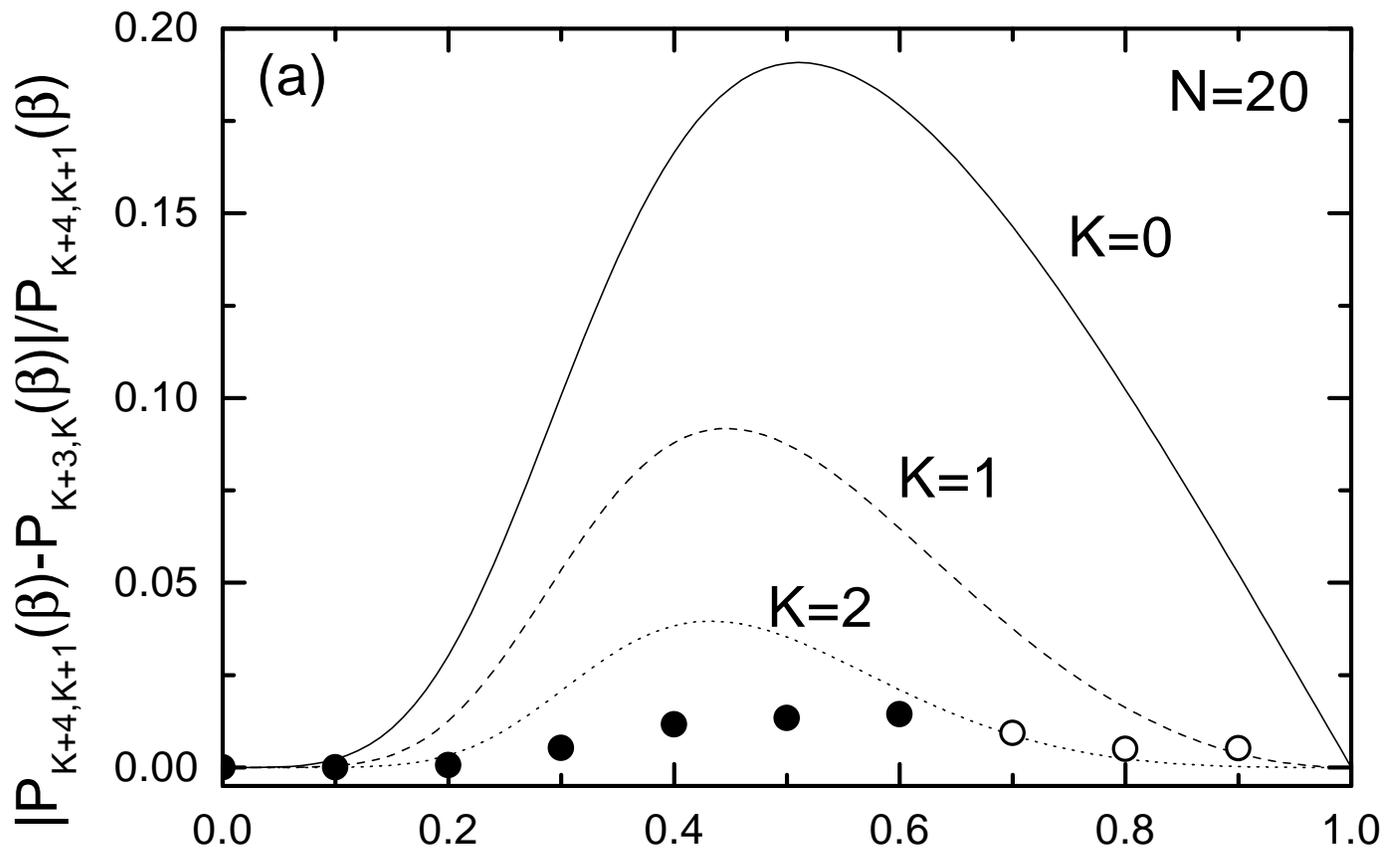
\begin{figure}
\caption{The relative differences between consecutive
Pad\'{e} approximants of the sequence $P_{K+3,K}(\beta)$. We consider 
two systems: (a) $N=20$ and (b) $N=210$. The dots give the relative 
difference $|E_{VMC}-P_{6,3}(\beta)|/P_{6,3}(\beta)$
between the best Pad\'{e} approximant and the variational Monte Carlo 
(VMC) results.}
\label{fig1}
\end{figure}

In Figs.~\ref{fig1}(a) and~\ref{fig1}(b) we show
the relative differences between consecutive 
approximants for 20 and 210 bosons, respectively. 
From these figures, the maximum error of the $P_{6,3}(\beta)$ approximant
is estimated to be lower than 4\% for any value of $\beta$. For $N=20
(210)$ 
the maximum error is $3 (4)\%$ which is reached for $\beta=0.8 (0.4)$. 

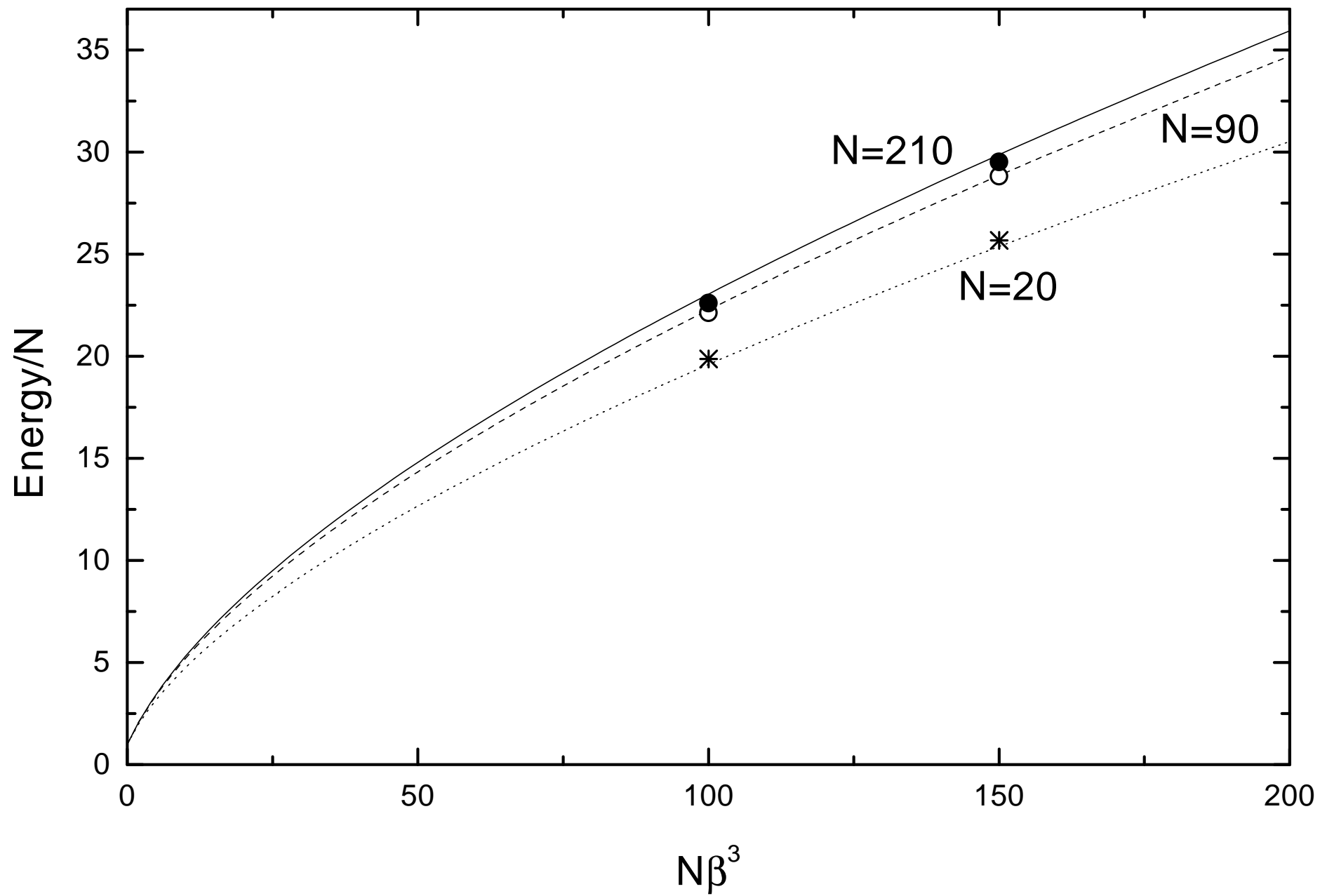
\begin{figure}
\caption{Approximate scaling properties of the energy for $N=90$ and 
$N=210$. As a comparison we also show the result for $N=20$. 
The dots (solid dots for $N=210$, open dots for $N=90$ and stars for
$N=20$) are the results from the variational Monte Carlo calculation.}
\label{fig2}
\end{figure}

The explicit form of the coefficients $a_k$ and $b_k$ suggests that 
$\epsilon/N$ is an ``almost universal'' function of the variable 
$N \beta^3$ for large $N$. We found that for
$N\ge 90$, the coefficients reach their asymptotic forms, and we 
have

\begin{eqnarray}
b_k \beta^k &\approx& N \tilde b_k (N \beta^3)^{k/3}, \\
a_k/\beta^k &\approx& N^{5/3} \tilde a_k/(N \beta^3)^{k/3},
\end{eqnarray}

\noindent
where $\tilde b_0=1$, $\tilde b_3=\frac{1}{2}\sqrt{\frac{\pi}{2}}$,
$\tilde
b_6=-0.05300695$, $\tilde a_0=1.062$ and $\tilde a_2=0.573 N^{1/4}$. The
coefficients $\tilde b_k$ and $\tilde a_k$ are numbers, except for
the coefficient $\tilde a_2$ which exhibits a smooth dependence on $N$. In
the interval $90\le N \le 210$ $\tilde a_2$ is approximately a constant
($\approx 2$) and consequently for the energy we obtain

\begin{equation}
\epsilon(\beta) \approx N f(N \beta^3),
\end{equation}

\noindent
where $f$ is only a function of $N \beta^3$. 
The $P_{6,3}(\beta)$ estimates for $N=20, N=90$  and $N=210$
are given in Fig.~\ref{fig2}, showing the ``approximate'' scaling when
$N\ge 90$. 
Note that bosons interacting through short-range potentials also show
scaling behaviour in the large-$N$ limit. Indeed, in this limit they are
described by the Gross-Pitaevskii equation, which may be written in a 
scaled form.\cite{GPS97a}

In order to have an independent check on the accuracy of the present
Pad\'{e} approximant, we carried out variational Monte Carlo (VMC)
calculations~\cite{MC} for the ground-state energy. The trial wave function 
used in the computations was (up to a normalisation constant)

\begin{equation}
\Psi_T=\left( \prod_{i=1}^N \phi(r_i)\right) e^{-\sum_{i<j} u(r_{ij})},
\end{equation}

\noindent
where $\phi(r)=e^{-r^2/2}$ is the ground-state function of a boson in 
the harmonic potential, and the pseudopotential $u(r_{ij})$ 
was chosen as

\begin{equation}
u(r) = \frac{\beta^3}{1+r^3}(A-r-\frac{C}{1+\beta} r^3 {\rm ln}~r),
\end{equation}

\noindent
where $A$ and $C$ are variational parameters. This 
interpolative expression for $u$ gives the correct asymptotic form 
at coincidence, $r\to 0$ (``cusp conditions''), 

\begin{equation}
\left. u(r)\right|_{r\to 0} = {\rm const} -\beta^3 r+\dots,
\end{equation}

\noindent
and at very large values of $r$,

\begin{equation}
\left. u(r)\right|_{r\to \infty} = -C \frac{\beta^3}{1+\beta} 
     {\rm ln}~r+\dots.
\end{equation}

\noindent
The later expression can be easily verified for the $N=2$ system. 
The coefficient in front of $-{\rm ln}~r$ is roughly the difference
$\epsilon(\beta)-\epsilon(0)$, which is proportional to $\beta^3$
for small $\beta$ values, and tends to $\beta^2$ as $\beta$  
increases. 

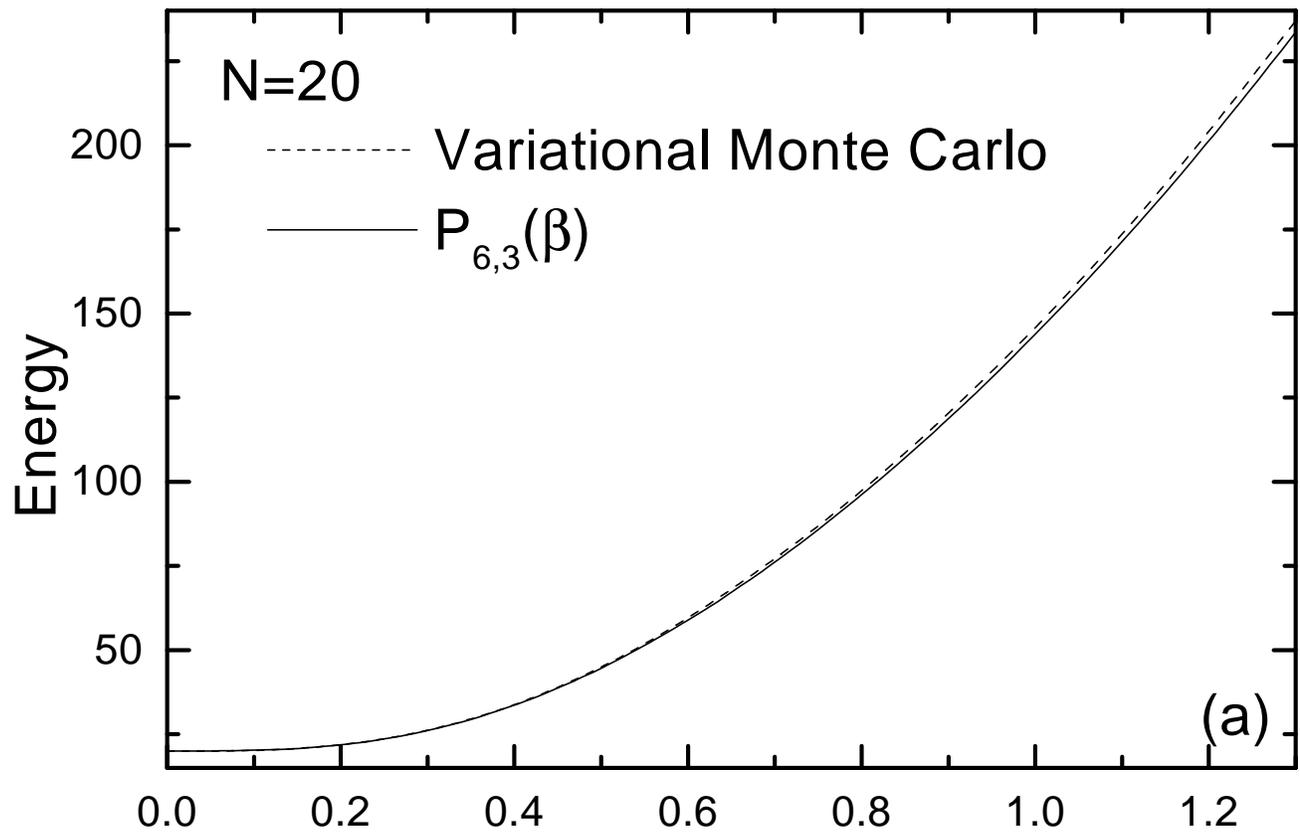
\begin{figure}
\caption{Comparison between  the $P_{6,3}(\beta)$ Pad\'e estimates (solid
curves) and the variational MC calculations (dashed curves) for: (a)
$N=20$ and (b) $N=210$.}
\label{fig3}
\end{figure}

Fig.~\ref{fig3} compares the Monte Carlo and $P_{6,3}(\beta)$
Pad\'e results for 20  and 
210 bosons as function of $\beta$ in the interval $0\le\beta\le 1.3$. In the 
computations, $10^5$ sweeps were used to compute mean values after 
$2\times 10^4$ steps for thermalisation. 
Note that $\beta=1.3$ is well outside the perturbative 
regime. Indeed, $\epsilon(1.3)/\epsilon(0)>12$ for $N=20$, whereas 
$\epsilon(1.3)/\epsilon(0)>35$ for $N=210$. 
\par
In Fig.~\ref{fig1} we have also plotted (the dots) the relative difference
between the $P_{6,3}(\beta)$ Pad\'{e} approximant and the variational Monte
Carlo results. Note that in Fig.~\ref{fig1}(a) for $N=20$ up 
to $\beta \simeq 2.5$ (the solid dots) 
the $P_{6,3}(\beta)$ approximant gives
practically identical results as those obtained from the variational Monte
Carlo. The relative error is less than $1.5 \%$ which is below the $4\%$ level
estimated previously from a comparison between successive Pad\'{e} approximants.
For larger values of $\beta$ one approaches the Wigner limit where the
particles form a crystal-like structure. In that limit
the functional form taken for the trial function is expected to be 
no longer good and we notice that
the relative error does not decrease with $\beta$ (open dots) but stays
approximately constant. Also for $N=210$ it is shown in Fig.~\ref{fig1}(b) that
the relative error is less than $4\%$.
\par
\acknowledgements
AG acknowledges support from the Colombian
Institute for Science and Technology (COLCIENCIAS). BP is an aspirant,
and FMP a research director with the Flemish Science Foundation
(FWO Vlaanderen). Part of this work is supported by FWO, a NATO Linkage grant, 
the `Interuniversity Poles of Attraction Programme - Belgian State, 
Prime Minister's Office - Federal Office
for Scientific, Technical and Cultural Affairs', and the WOG on field theory
and statistical physics.

\appendix

\section*{Second order energy correction for the two particle
interaction problem}

Here we present an alternative but more accurate calculation of the coefficient $c_2$
for the second order energy correction (\ref{second}). 
Because expression (\ref{second}) exhibits a rather simple
dependence on the number of bosons $N$ we can limit ourselves to calculate  
the coefficient $c_2$ 
for the system of two particles with Coulomb repulsion in a parabolic
confinement potential. The ground-state of that system will be
symmetric with respect to the permutation of the particles what automatically
takes into account the boson character of the problem.

It is known that in this two particle interaction problem 
the relative and the center-of-mass motion can be separated. 
The center-of-mass
($\vec{R}=(\vec{r}_1+\vec{r}_2)/2$) motion is the one of a harmonic oscillator
and is easily eliminated. Consequently, we have only to consider the
following radial Schr\"{o}dinger equation for the relative motion
($\vec{r}=\vec{r}_1-\vec{r}_2$) of particles
\begin{eqnarray}
  \left\{H_0+\lambda/r-E\right\}R(r) = 0, \\
  \quad H_0 = -\frac{1}{r}\frac{d}{dr}r\frac{d}{dr} +\frac{1}{4}r^2,
\end{eqnarray}
with $\lambda=\beta^3$.

Now let us expand the eigenvalue $E=E_0+\lambda
E_1+\cdots$ 
and the wave function $R=R_0+\lambda R_1+\cdots$ into powers of $\lambda$.
The application of the standard
perturbation technique leads to the following expressions
\begin{eqnarray}
  E_0 &=& 1, \\
  E_1 &=& \int_0^{\infty}drR_0^2(r) = \sqrt{\pi/2}, \\
  E_2 &=& \int_0^{\infty}dr R_0(r)R_1(r),\label{energy2}
\end{eqnarray}
with $R_0 = \exp(-r^2/4)$ and
the first order wave function correction obeys the following equation
\begin{equation}\label{first}
  \left\{H_0-E_0\right\}R_1 = (1/r-E_1)R_0.
\end{equation}
This equation has to be solved together with the boundary and orthogonality
conditions
\begin{equation}\label{cond}
  R_1(0)=R_1(\infty)=0, \quad \int_0^{\infty}drrR_1(r)R_0(r) = 0.
\end{equation}
Inserting the substitution $R_1 = \{\sum_{n=1}^{\infty}w_nr^n+C\}R_0$
into equation (\ref{first}) we obtain the recurrence relation
\begin{equation}
  w_{n+2} = w_n n/(n+2)^2, \quad w_1=1, \quad w_2=-E_1/4.
\end{equation}
Now iterating the above relation, withy the ortogonality condition
(\ref{cond}), and inserting the obtained result into expression (\ref{energy2})
we obtain the second order energy correction expressed as a single sum of
$\Gamma$-functions
\begin{eqnarray}
  E_2 &=&  \sqrt{\frac{\pi}{16}}\sum_{k=0}^{\infty} \left\{
  \frac{\Gamma(k+1)}{(k+1/2)\Gamma(k+3/2)} -
  \frac{\Gamma(k+3/2)}{(k+1)\Gamma(k+2)} \right\} \nonumber\\
  &-& \frac{\pi}{8} \sum_{k=0}^{\infty}\frac{1}{(k+1/2)(k+1)}.
  \label{laatste}
\end{eqnarray}
The coefficient $c_2=E_2/2-c_1$ follows from expression (\ref{second}) with
$N=2$. The sums in expression~(\ref{laatste}) converge which enables us to
obtain the coefficient $c_2$ up to the desired accuracy.

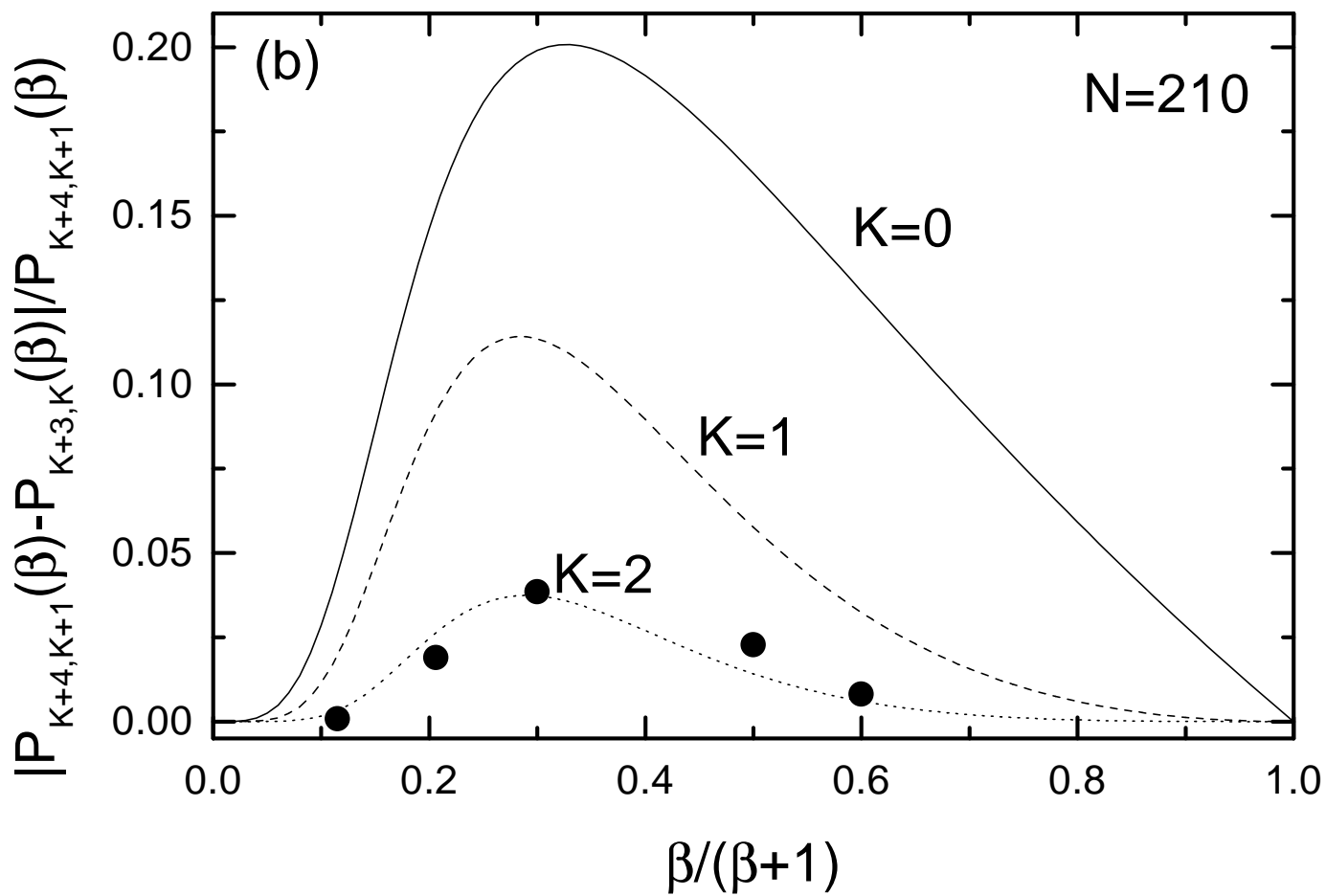
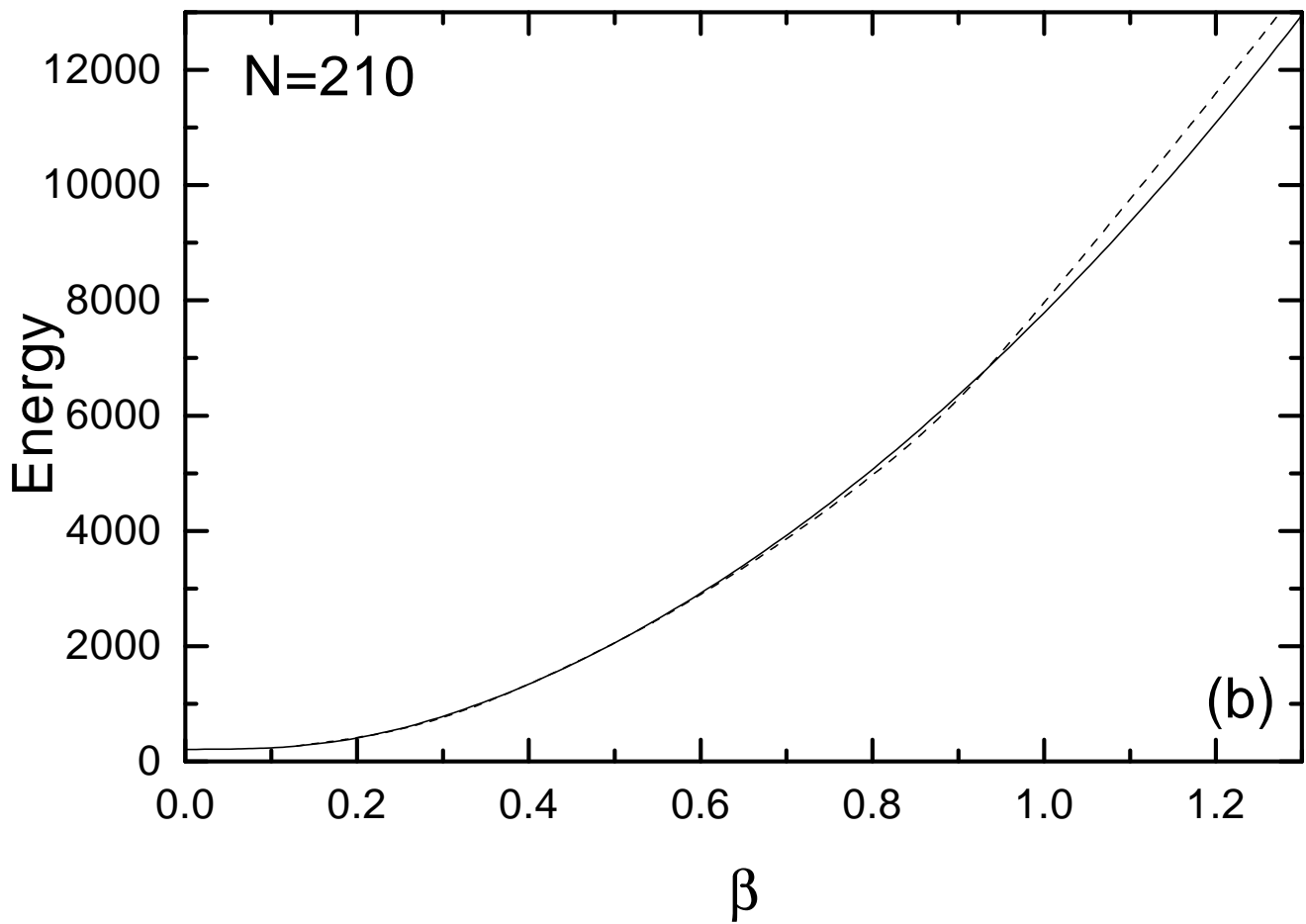
\end{document}